# Developing Concurrent Coding: An unconventional encoding scheme applied to visible light communications

David M. Benton[a]

a Aston Institute of Photonic Technologies, Aston University, Aston Triangle, Birmingham B4 7ET

**Abstract**. An unconventional encoding scheme called concurrent coding, has recently been demonstrated and shown to offer interesting features and benefits in comparison to conventional techniques, e.g. robustness against burst errors and improved efficiency of transmitted power. This concept has been demonstrated for the first time with optical communications where standard light emitting diodes (LEDs) have been used to transmit information encoded with concurrent coding. The technique successfully transmits and decodes data despite unpredictable interruptions to the transmission causing significant drop-outs to the detected signal. The technique also shows how it is possible to send a single block of data in isolation with no pre-synchronisation required between transmitter and receiver, and no specific synchronisation sequence appended to the transmission. This work also demonstrates for the first time the successful use of multithreaded (overlaid) concurrent codes.

**Keywords**: Encoding, free space optics, visible light communication

E-mail: d.benton@aston.ac.uk

## Introduction

Concurrent coding is a unique method of encoding that differs significantly from conventional encoding methods[[1]-[5]] and has recently been investigated as a novel, robust method of protecting data transmission[[6]]. Conventional approaches to protecting information transfer against the corrupting effects of noise see the characteristics of a block of information being encoded with and into the information, thus increasing the amount of data being sent and reducing the information rate. This characteristic information is almost always linked locally to the information itself, such as a parity bit next to the data bits it represents. This is true for block codes (see[[7],[8]]) including cyclic codes, Golay codes, Reed Solomon codes[[9]] and is equally true for convolutional codes such as turbo coding[[10]] and Viterbi codes[[11]]. These schemes are designed to combat the effects of random noise affecting individual isolated bits and do not deal effectively with non-random errors [[12],[13]]. In order to combat the effect of burst errors which affect a contiguous set of bits in a non-random way, interleaving is typically used. This deliberately converts the local connection between related coded bits by distributing them in a regular fashion throughout a larger codeword. Thus random errors and burst errors are treated separately. Alternative approaches such as Fire codes[[14]] and Reed Solomon[ [9],[15]] encoding treat data as a set of symbols and correct for symbol errors to help encompass non-random errors. Concurrent coding connects the characteristics of a block of data to the codeword in which it is transmitted. The data block is encoded globally into the codeword in a single step.

Concurrent coding is a binary asymmetric technique that encodes and decodes message vectors rather than bits but can only be implemented on binary modulation schemes. It has shown robustness against noise and in particular burst errors[[6]]. It was originally conceived as a method for providing protection against jamming, an alternative to spread spectrum techniques [[1],[2],[3]] . Concurrent coding can achieve jamming resistance without the need for a shared secret key as is required with code division multiple access (CDMA) coding. Concurrent coding works by repeated use of a hashing function to distribute information throughout a codeword. Many hashing functions are appropriate [[4],[6],[16]] but the emphasis is on distribution rather than security although attacks against the algorithm have been examined[[17]] . Recently Hanifi *et al* [[18]] have developed a new concurrent code based on the use of monotone Boolean functions.

Concurrent codes are appealing for their robust nature but also for other properties such as the efficient use of transmitted energy and the relative simplicity of the scheme in comparison to other comparable techniques such as Reed Solomon encoding. Concurrent codes degrade more gracefully than interleaved codes where data loss increases dramatically once the error fraction increases beyond the code's capacity to correct it[[19]]. With concurrent codes data is (ideally) not lost, only obscured. Thus concurrent coding could be a suitable protocol to apply to free space optical connections where burst errors are a particular problem due to beam interruptions, misalignments and atmospheric scintillation. On-Off keying (OOK) is a commonly used intensity modulation scheme used in optical communications [[20],[21],[22]] in which a binary representation is obtained from the presence or absence of light – hence optical communication is a natural ally for concurrent coding. Used with direct detection OOK requires a knowledge of the instantaneous fading coefficient of the channel in order for dynamic thresholding to be applied. In this sense, using concurrent coding with OOK, the encoding scheme is the modulation scheme. Other modulation schemes such as pulse position modulation (PPM), which are typically symbol transmission schemes, are not compatible with concurrent coding unless implemented in an asymmetric binary manner (i.e. large slot for binary 1) which would reduce their efficiency. The effect of atmospheric turbulence in free space optical (FSO) links can result in very large and deep signal fades. No matter which encoding scheme is used, the need for large scale interleaving has been required for the successful operation of the encoding scheme. Concurrent coding may be the first genuine alternative to interleaving in FSO communication systems.

Comparison of the behaviour of the encoding methods is therefore appropriate. However it is important to first highlight the nature of concurrent codes to appreciate the difference with conventional codes and to get a proper appreciation of the comparative behaviours.

Concurrent codes are an asymmetric binary encoding system that generates indelible marks to represent digital 1's into a codeword. Marks are substantive; a positive presence such as pulses of energy and cannot be removed to randomly convert a 1 back to a 0. A zero is then the absence of energy or substance which can be converted to a 1 by noise (or jamming). A result of using indelible marks is that encoded message vectors cannot be removed and will always be decoded.

Original message vectors cannot be corrupted but can be obscured by spurious decodings called hallucinations.

Providing protection for transmitted data against random errors, burst errors and jamming might involve separate steps for each error and could be represented as

Data → Parity encoding → Interleaving → Spread spectrum coding → Transmission.

In contrast concurrent coding follows;

Data → Concurrent coding → transmission

Marks in the codeword are shared by many input vectors thus leading to improved efficiency in terms of transmitted energy.

Concurrent codes contain an inherent method of synchronisation

Concurrent codes are significantly easier to comprehend than say Reed Solomon encoding or other block codes.

## Description of Concurrent Coding

Descriptions of how concurrent coding works are given in previous references[[1]-[6]] and briefly given here for completeness. The concurrent coding principle encodes a digital word into a much larger codeword space by hashing incrementally increasing sub-sections (or prefixes) of the digital word to produce addresses in the larger codeword, into which indelible marks are placed to represent 1's. Thus a 4 bit message vector *abcd* would place marks in the codeword at positions given by *H(d), H(cd), H(bcd)* and *H(abcd)* where *H(x)* represents the output of hash function upon the digital sub-vector *x*. This is represented schematically in the decoding section of Figure 1. The hash function is not explicitly defined and can be any suitable redistribution function. The process is repeated for additional message vectors again placing marks into the codeword. When all message vectors are installed the codeword can be transmitted. Decoding the received codeword proceeds as follows: The receiver computes the mark positions for *H(0)* and *H(1)* and then checks the codeword to see if any of these marks are present - so *H(d)* would correspond to one of these. These mark positions form the first branches of a decoding tree and where corresponding marks are found live branches are recorded. If specific marks are not found all branches stemming from that point cannot exist in the codeword and these dead branches are not investigated further. Decoding then proceeds by calculating (assuming both *H(0)* and *H(1)* are live) marks positions for *H(00), H(10), H(01)* and *H(11)*. Again where corresponding marks are found live branches are retained. This process repeats with the number of decoding rounds equal to the length of the message vectors. At the end of decoding the remaining message vectors represent the original message vectors. This is shown schematically in the decoding portion of Figure 1. If the codeword is subject to large amounts of noise then it is possible for spurious messages to get through the decoding tree and these are termed hallucinations. Note that the original message vectors will always be decoded – a result of using indelible marks – but the effect of hallucinations will be to obscure which messages are genuine. In order to kill hallucinations additional checksum bits with fixed values are added to the message vectors so that for the earlier example *11abcd* would be encoded. This increases the number of decoding rounds but is very effective at killing off hallucinations. Burst errors - such as arise from obscurations or blockages of the signal, appear as blocks of contiguous zeros in the codeword. When these 'gaps' in the codeword become large enough to be statistically unlikely to occur by chance, the decoding round can keep alive any hash calculations that result in marks that would appear in the gap. Allowing decoding branches to connect across the gap is an effective way of correcting for burst errors where even large missing fractions of the codeword (up to 40%) can still result in perfect decoding. Once again all genuine message vectors are decoded with hallucinations appearing. A gap is made statistically significant by ensuring a minimum number of message vectors are encoded so that the randomising effect of the hash function fills the codeword evenly with marks.

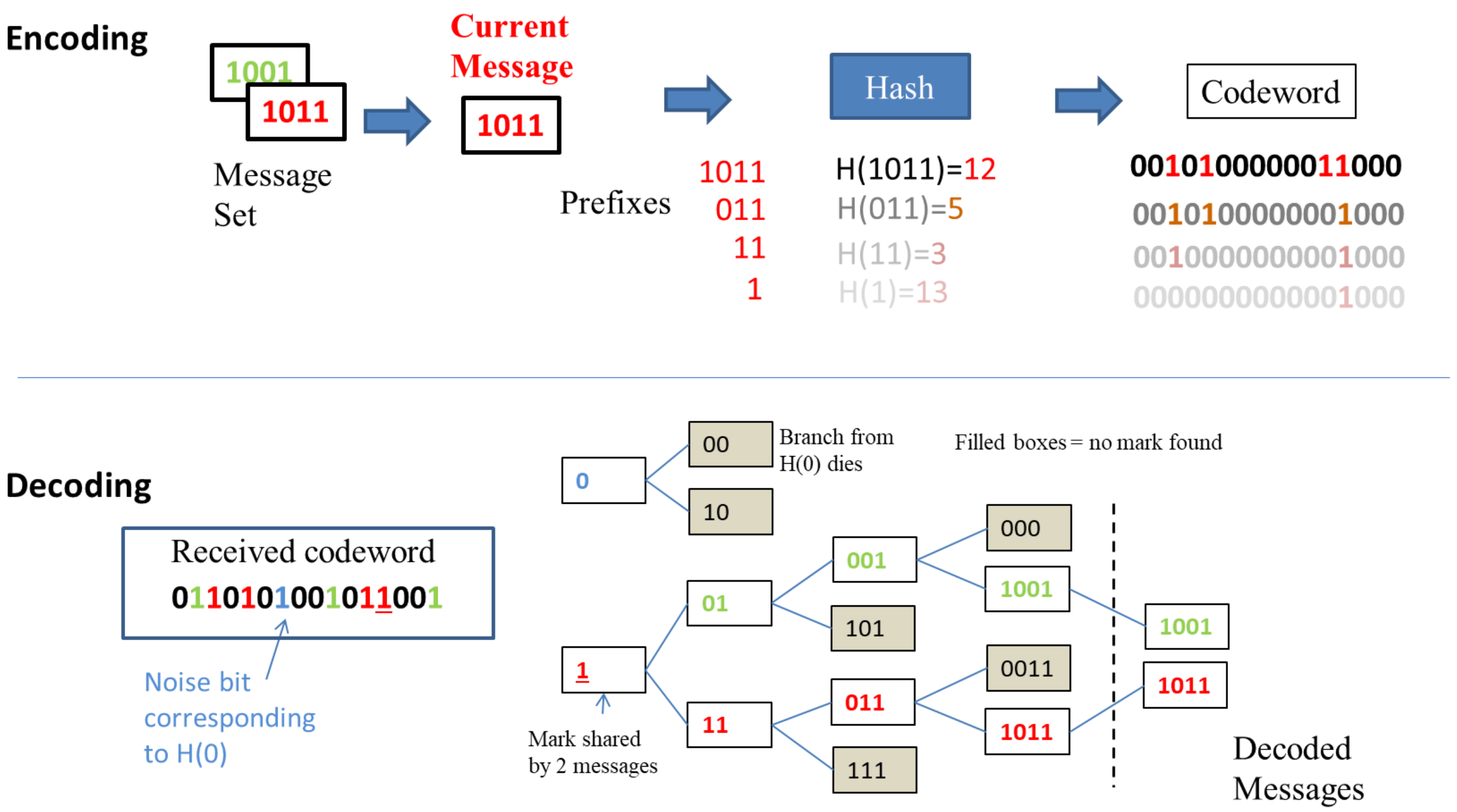

**Figure 1. A schematic representation of the encoding process (top) and the decoding tree (bottom). Prefixes of the input message are hashed to produce the addresses of marks in the codeword. The codeword is examined by comparing possible prefix hashes with the presence of marks in the codeword.**

It is the robustness of this single encoding step that is of relevance to FSO links, where performance against burst errors is better than for pure interleaving. Burst errors due to atmospheric effects or unstable alignment are a major issue in the implementation of FSO links. In this work topics are addressed that make the implementation of concurrent coding achievable and practical. It shows that multiple encoding streams can be overlaid into a single transmission, similar to the way multiple users can be overlaid with CDMA encoding. Also demonstrated is a property of concurrent coding that allows any single transmission to be decoded through the use of an inherent synchronisation property. Finally the first use of current coding with a visible light channel is demonstrated with multiple threads and burst errors present.  This work was first presented in a conference proceedings[[26]]

*Modelling of Parallel Hashes*

Previous work has highlighted the fact that concurrent coding produces fewer marks (1's) in its codeword than other techniques[[6]], leading to a reduced transmission power. In comparison to Code Division Multiple Access (CDMA) encoding the transmitted power can be orders of magnitude less [[19]]. The efficiency of concurrent coding arises from common sub-sequences sharing marks in the codeword. This efficiency comes at the cost of being able to decode any particular message only once, as multiple encoding will simply reproduce the same marks. There is also no specific order to the decoded messages which are all decoded in parallel. This might not be an issue in situations such as a sensor network where individual sensors include a sensor id with their data transmission. However for more general communication more information will be needed. Reference was made in [[6]] to the use of multiple hash functions for the purpose of encoding the same information more than once, or for providing decoding guidance information. In this work the use of multiple hash functions within the same codeword is investigated and implemented.

The hash function used for concurrent coding can be any generic function that can redistribute data patterns throughout a large codeword in a suitably dispersed fashion. Various redistribution functions have been used as hash functions including a pseudo-random bit sequence (PRBS), Glowworm hash [[16],[17]] and the FNV hash[23] etc. To ensure no collisions in the hash function we use hash tables for a so-called *closed* concurrent code [[19]] where output addresses have been randomly and uniquely defined. Hash tables are not the ideal method for practical implementation- particularly with larger codewords – but here they suit our requirements.

Building upon previous models, 8-bit messages with 2 checksum bits were encoded into a codeword space of $2^{11}$ bits. A set of 2048 element hash tables was generated and each table assigned a thread number. A set of random messages were selected and then encoded within each thread into a single codeword – this will show that the same information can be encoded multiple times. Another method for overlaying data is to use a single hash function with a cyclic positional offset added, with a different offset representing each thread. In principle the hash functions could all be completely different such as a mixture of PRBS, glowworm, hash table etc.

The number of marks produced for a given number of messages m, is given by

$$Z(m) = (N + k)m - m log_2 m + \frac{3m}{2} \qquad (1)$$

where *N* is the number of bits in a message and *k* is the number of checksum bits. This nonlinear increase in the number of marks is caused by the sharing of marks by multiple messages and is the source of the efficiency of concurrent coding. This is true for a single thread (i.e. data encoded with a particular hash function) however multiple threads are independent. Because concurrent coding is an OR channel independent threads can also share marks. For a small number of messages and a few threads we would not expect this occur often, but with increasing messages and threads this will become more prevalent. A simulation involving a concurrent code with multiple hash functions was created to investigate the properties and behavior. This can be seen in Figure 2 where the actual number of marks produced by a concurrent code is plotted for variation in the number of encoded messages and with different number of encoded threads with each containing the same messages. For a small number of threads the number of marks increases linearly with threads, but as the number of threads is increased the relative increase in the number of marks is reduced. This is an indication that different threads are sharing marks in the codeword.

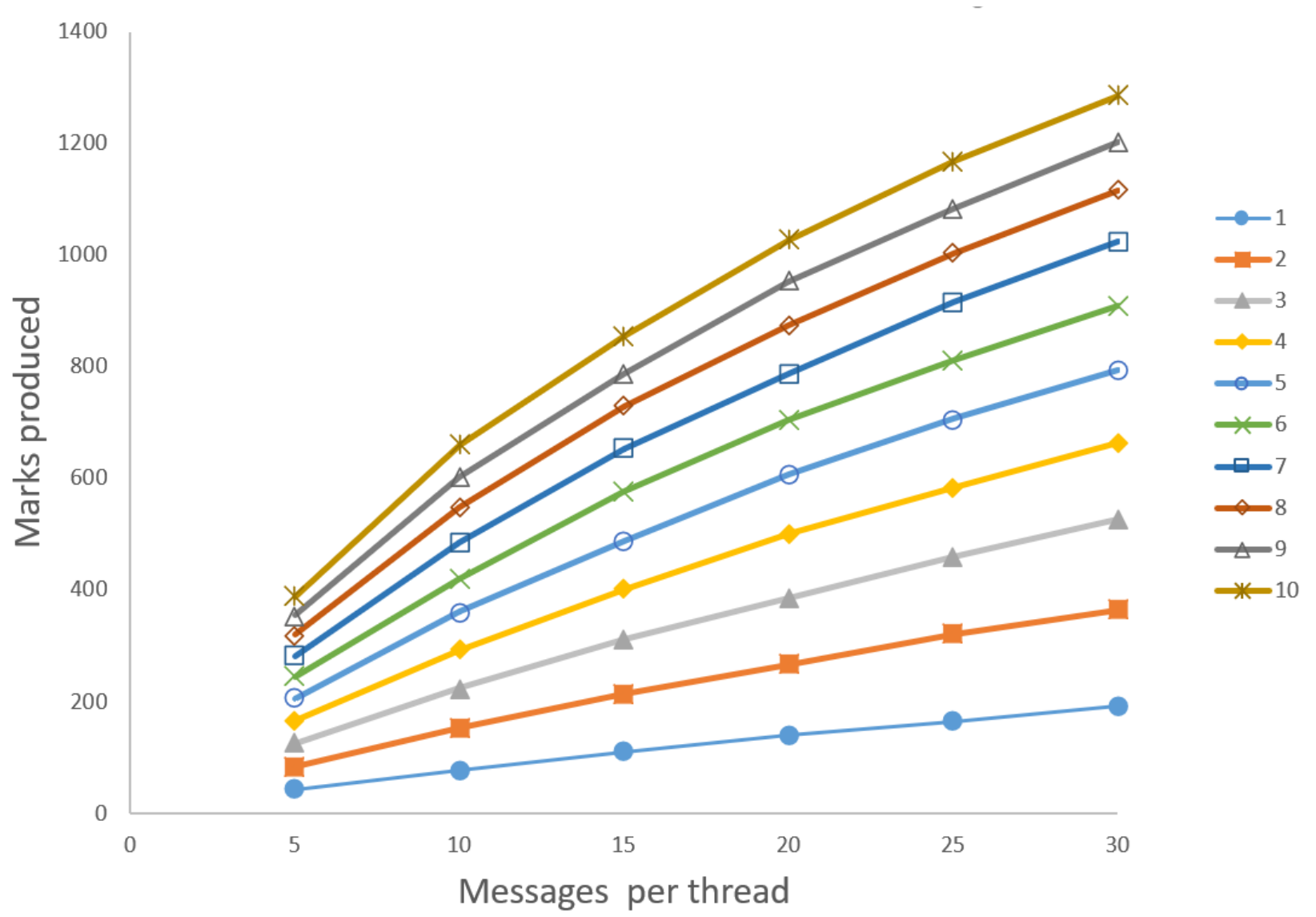

**Figure 2. The total number of marks varying with number of encoded messages and number of threads. Symbols represent the measured number of marks produced in the code-decode process by a computer simulation running a concurrent code algorithm. Solid lines are the expected values from modelling**.

Because the threads are independent, any one thread sees marks from other threads as noise and this can influence the production of spurious false decodings (hallucinations) across all threads. We can model the number of marks as follows: Each thread is added serially into the codeword. Thus each mark in the current thread has a probability of being shared with previous threads given by:

$$P = \frac{M_{i-1}}{C} \quad (2)$$

Where $M_{i-1}$ is the total number of marks produced by all previous threads and C is the codeword length. The number of shared marks produced is then

$$S_i = \frac{Z_i \, M_{i-1}}{C} \quad (3)$$

Where $Z_i$ is the number of marks produced by the current independent thread according to equation 1. The total number of marks produced after the *ith* thread is added is:

$$M_i = M_{i-1} + Z_i - S_i \quad (4)$$

This iterative relation was used to calculate the expected number of marks shown by the solid lines in Figure 2 and in good agreement with the data points. Figure 3 shows the number of measured hallucinations produced in an actual concurrent code (using parameters given earlier), as the number of threads and the number of messages per thread is increased. More hallucinations are produced for the same number of marks when the number of threads is increased. Figure 4 is a bubble chart plot showing the production of hallucinations as the number of threads and the number of messages per thread is increased. This is instructive in showing us the boundaries at which multi-threading challenges the encoding integrity. As a rough guide when the total number of marks exceed 1/4 of the codeword size hallucinations will appear.

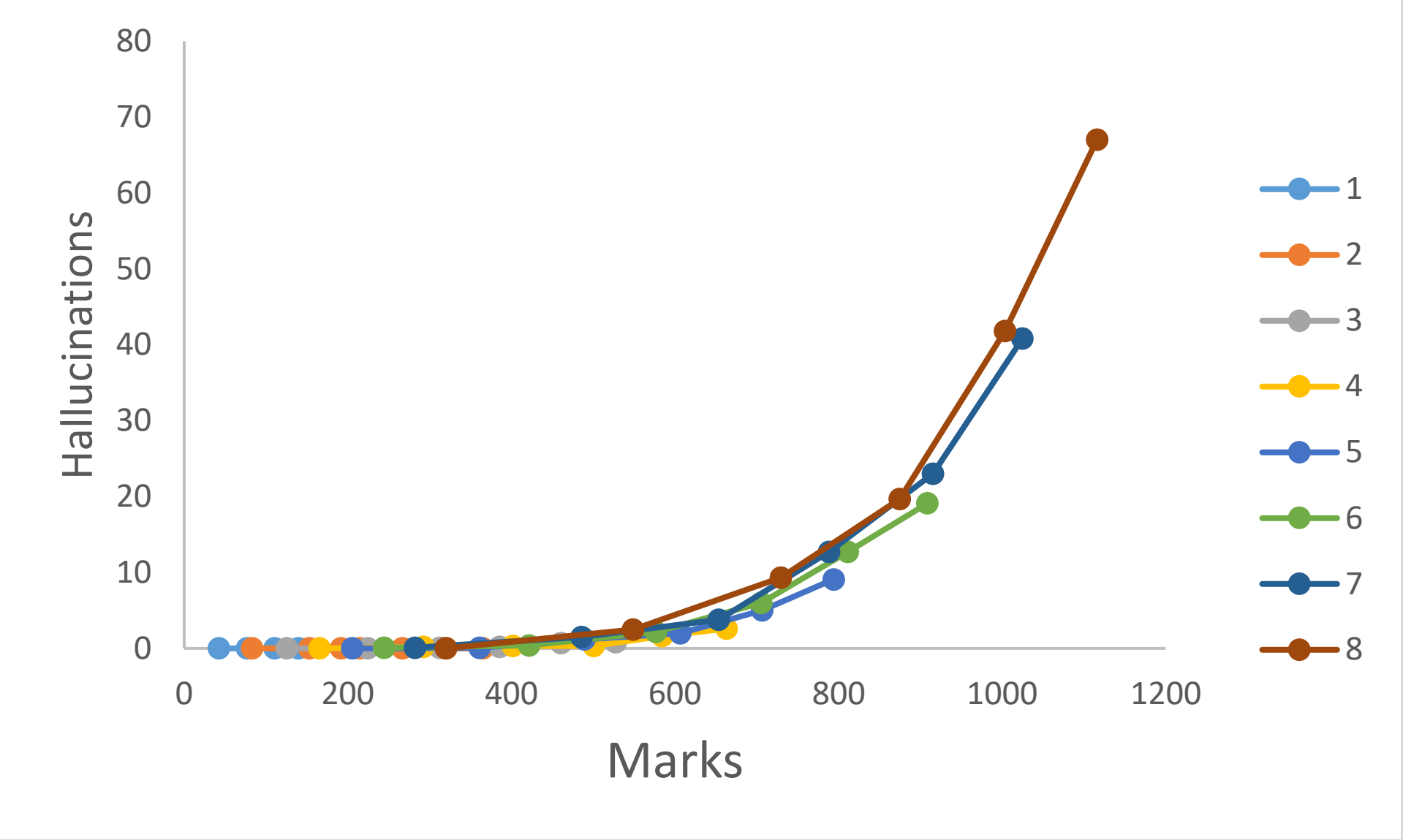

**Figure 3. The number of Hallucinations vs total marks generated for a limited number of threads, representing the data given in figure 2.**

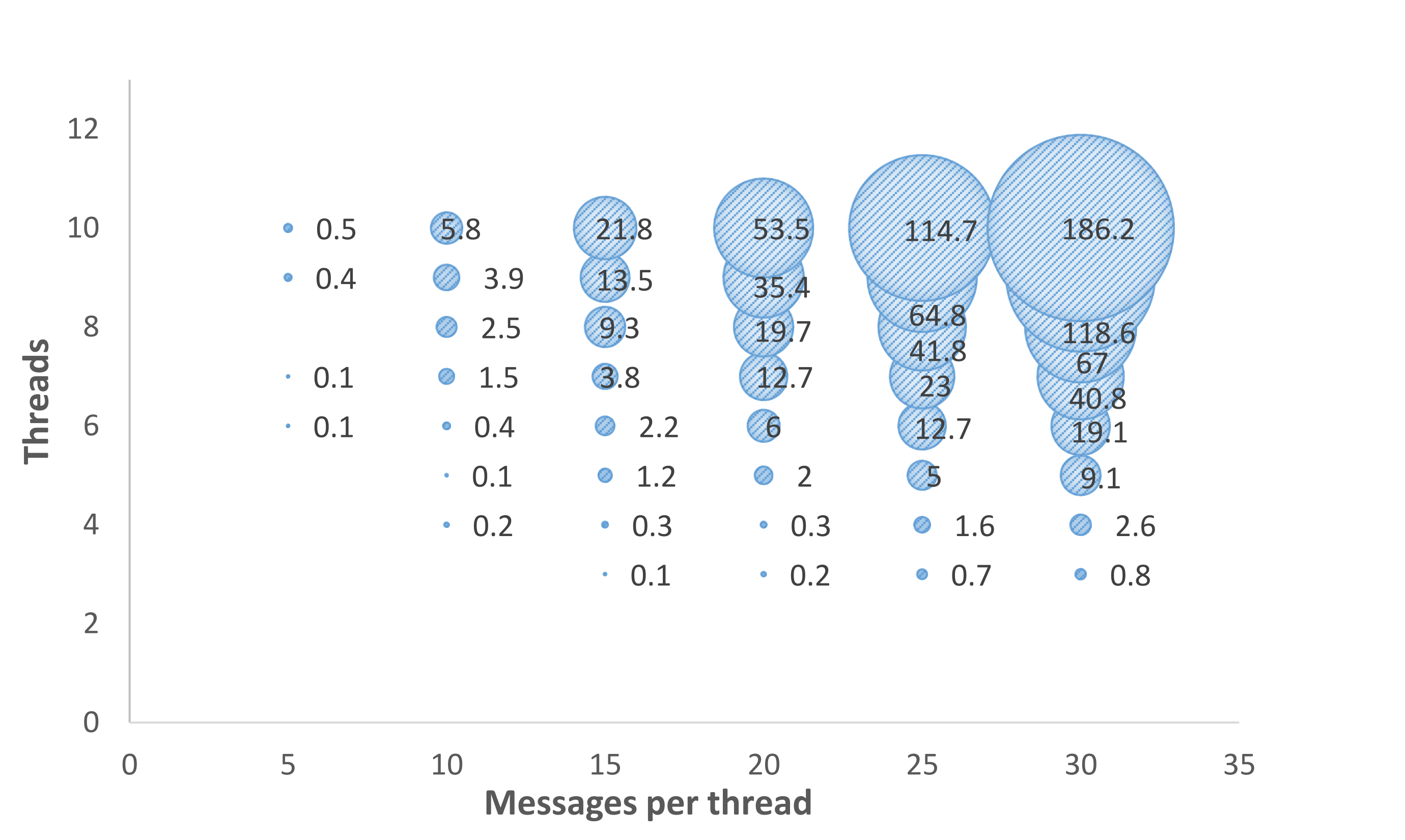

**Figure 4. A bubble chart showing the hallucinations produced as the number of threads and the messages per thread are varied.**

## Implementation

To demonstrate multithreaded concurrent coding a free space optical system was used consisting of 4 LEDs and a single photodiode detector. Individual threads were generated from hash tables using a LabView program. Each individual thread was sent to an LED using a USB Ni-DAQ data acquisition device and marks were represented by a light pulse in an OOK scheme. Each LED transmitted its codeword thread in parallel. A single silicon photodiode was located 60cm from the LEDs which were about 1cm apart and utilized their natural divergence to overlap their light onto the photodiode. The overlaid LED signals (an OR process) are then indistinguishable and were amplified and sent to a second computer via another USB Ni-DAQ unit which interpreted the pulses as digital signals rather than collecting analogue voltages – this prevented the system from using relative intensity to identify the emitting source. The requirement for each thread is that it uses a different hash function. This can be achieved in a number of ways such as using a different mathematical function in each case, or the same function with a different seed. In this work a set of hash tables common to both receiver and transmitter were used to ensure minimal interactions between threads.

Initial developments on synchronisation methods were performed with a single thread and LED. Multiple threads can of course be combined and transmitted by a single LED. Using multiple LEDs aligns with a multi-user system similar to that for CDMA techniques. In order to emulate the effects of burst errors that could be introduced by obstructions or beam wander, a galvanometer mirror was used to occasionally deflect the light way from the detector. This system is shown schematically in Figure 5

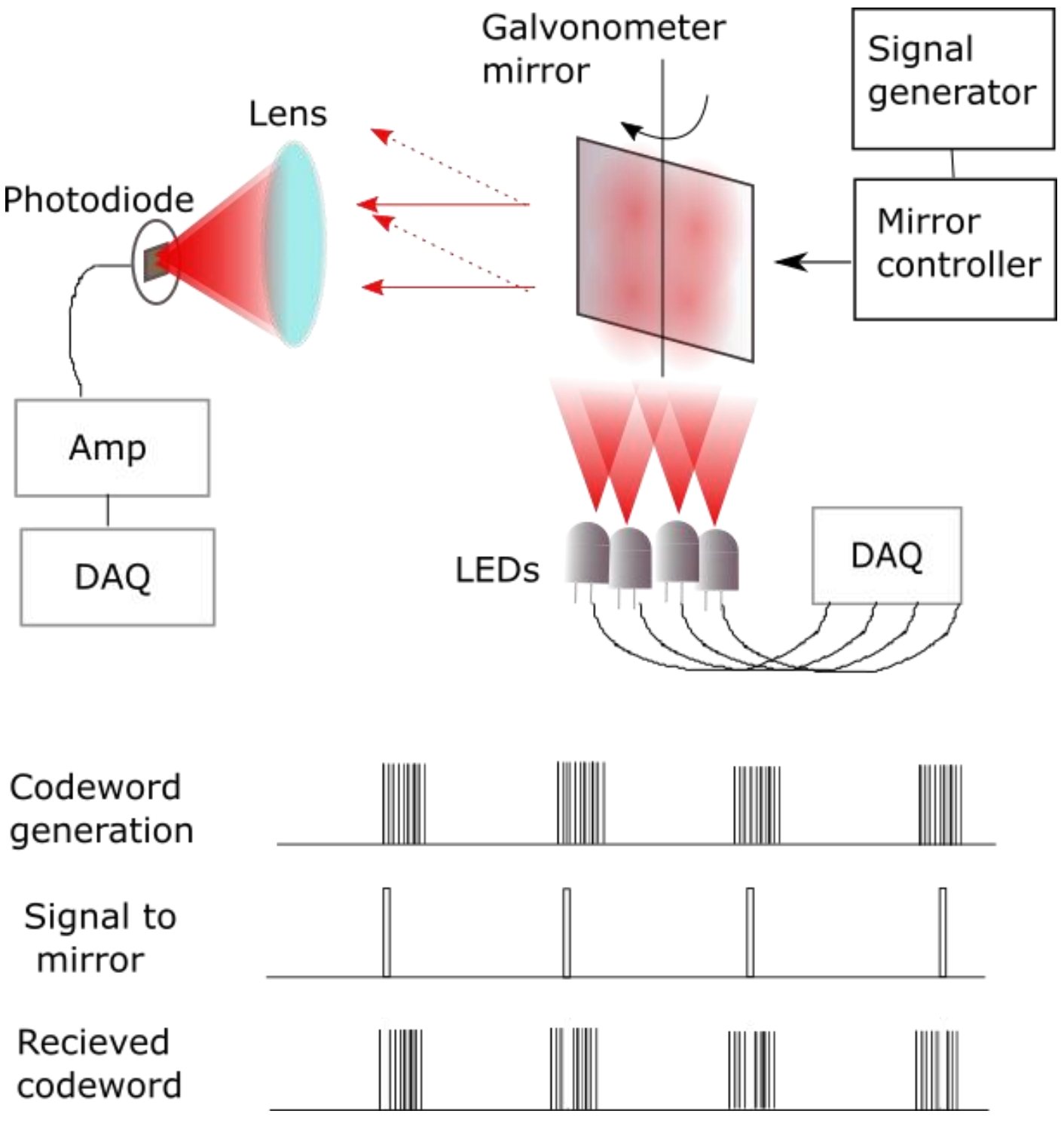

**Figure 5 A schematic diagram of 4 LEDs illuminating a photodiode detector via a galvanometer mirror (top). The generation of gaps in the received codeword is represented in the lower diagram.**

### *Synchronisation*

Hashed messages share marks that represent the first round of prefix encoding, thus the marks representing these prefixes will be occupied with high probability. The first two layers of prefixes - *H(0), H(1) and H(00), H(01), H(10), H(11)* - are occupied with high probability when a few messages are encoded and these marks – referred to as the principle marks - can be used to synchronise the received codeword [[19]].

The codewords were sent in isolation (that is with no pre or post information for synchronisation) with the intention that only the codeword itself can be used for decoding. This means that no prior setup to establish a phase locked loop was used. Before decoding can commence the boundaries of the codeword need to be established and the mark positions set. Because zeros are represented by the absence of signal the exact start of the codeword is not clear. This is established by setting up a synchronisation vector which is a codeword block containing only the principle marks (those corresponding to the 2 least significant bits in the decoding tree) from the first thread. This synchronisation vector is correlated with the received vector and the maximum correlation value taken as the reference point within the codeword from where the first mark position can then be determined. Note that this method becomes problematic for the use of a single hash function with multiple cyclic offsets as it produces multiple genuine correlation peaks and an additional step of identifying which peak corresponds to which thread would be required.

The transmitter and receiver were set to nominally the same sample frequency – typically 20kHz, however small differences of a few Hz lead to disparity in mark positions between receiver and transmitter. Across the length of the codeword a small drift causes marks to be spread across 2 mark positions or shifted by an entire mark. In this case misplaced marks lead to a failure to correctly decode the full contents of transmission. This was understood and pointed out by Bahn [[24],[25]] who quantifed the precision with which oscillators should be matched and suggested the use of ‘Bookend marks’ to define the start and end of the codeword. As has been stated for the ideal modelling case, the nature of indelible marks means that encoded messages cannot be removed and will always be decoded. However correct synchronisation is the essential property required to make this true.

Correcting for mark drift proceeds as follows: A window around the position of each of the principle marks was established, typically 3 bins wide. Marks within the received codeword that fall within these windows are identified as being the principle marks and their positions recorded. The received positions are plotted against the expected positions to generate a linear relationship. Using the gradient and offset of this relationship all marks in the codeword can then be adjusted to ensure there is a mark in the correct position within the codeword. Decoding can then proceed.

The transmitter would encode a fixed number of randomly selected messages. These messages were repeated in each thread in order to demonstrate how the same information can be encoded more than once. We can evaluate the decoding process by recording the number of decoded messages in each thread. The inherent synchronisation approach was observed to work and allow the 4 encoded threads to be successfully decoded. However, it would suffer from a weakness arising from a reliance on randomly filled principle marks. A threshold level for correlation was set, typically corresponding to matching 5 of the possible 6 principle marks. Occasionally this threshold is not passed and this results in no decoding at all. To overcome this issue 2 amendments were investigated. The first involves adding to the beginning of the codeword a 16-bit code that can be identified by correlation. The second involves distributing throughout the codeword a small number of static synchronisation marks (similar to the bookend marks[[25]] but not at the ends of the codeword) . The first approach is simple and adds to the codeword length a little. However this represents an easily identifiable weakness to a would-be jammer and a small corruption of this sequence would result in no decoding results (as would be true for bookend marks). This is equally true for burst errors which occur over the sequence. The second approach should be more robust against both jamming and burst errors due to distributing marks throughout the codeword. Both approaches were investigated in relation to their robustness against burst errors by artificially introducing gaps of missing data into the codeword prior to transmission. The gaps were added to a fixed point centrally in the codeword (so as not to completely corrupt the decoding). The number of decoded messages for 300 transmitted and decoded streams, each containing 10 messages from just a single thread with increasing gap sizes were determined.

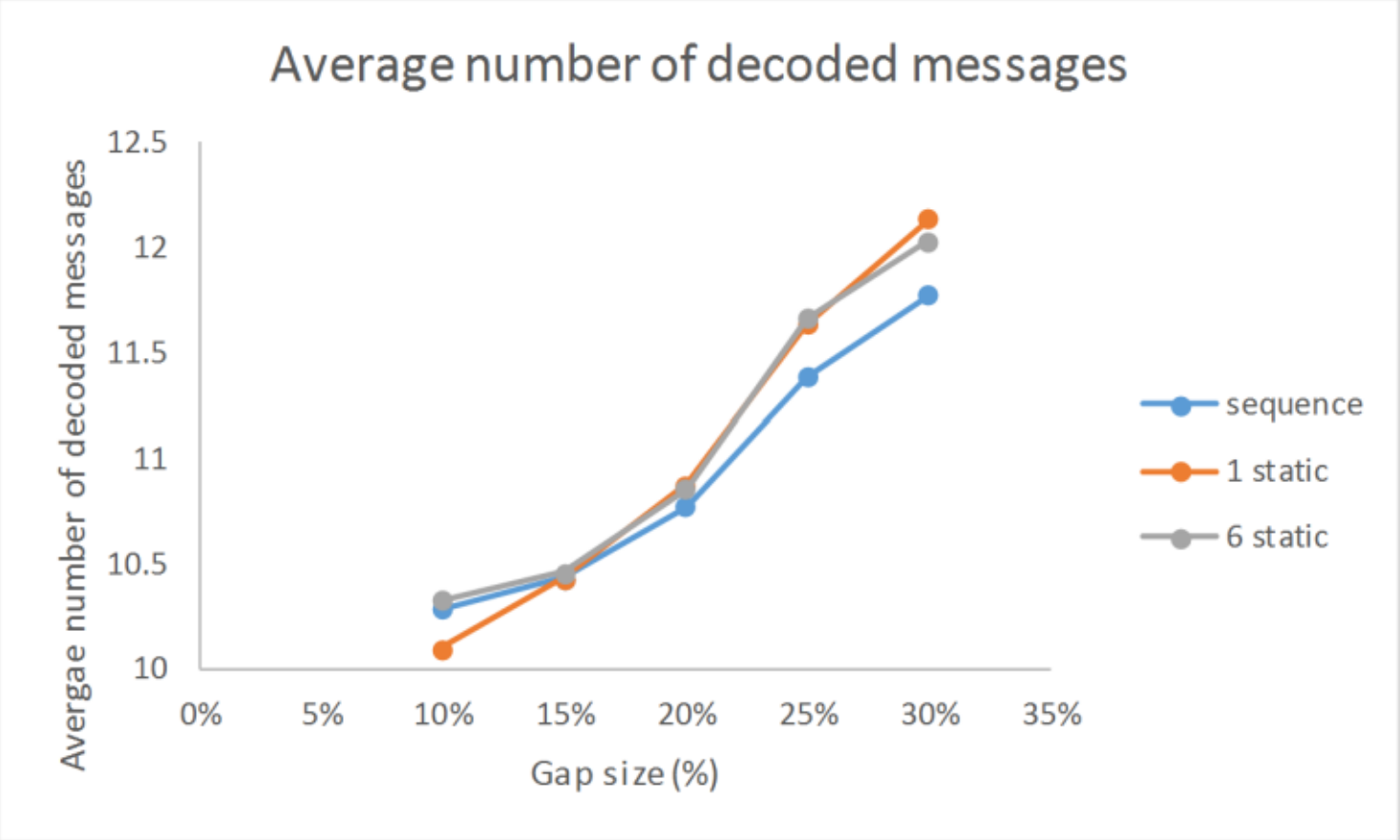

**Figure 6. The number of decoded messages for different synchronisation methods- an initial synch sequence, 1 additional static point (no sequence) and 6 additional static points.**

The plot in Figure 6 shows the average number of decoded messages against gap size for codewords involving a sequence, 1 static synchronisation point and 6 static points. It is clear from these data that hallucinations are being produced more significantly than was the case in modelling. This is most likely an effect caused by the synchronisation and especially the drift correction which would add marks into the codeword which would appear as noise. As was discussed in [[6]] concurrent codes have resilience to a combined level of burst error and noise. Further development of efficient and effective synchronisation schemes is required. It should be borne in mind that even though hallucinations are being produced, the original genuine messages are still being decoded with 30% of the codeword missing, which could not be done with interleaving.

In real world optical applications burst errors in the form of signal dropouts are unpredictable and can arise due to conditions such as misalignment or atmospheric scintillation. To better emulate this a galvanometer mounted mirror was used to steer the output light towards the detector. A square voltage pulse was applied to the mirror to cause it to twitch and briefly steer the light away from the detector. The pulse duration was chosen to ensure that the codeword gap was around 10% and the pulse was repeated at a rate such that the gap appeared in different places during the production of codewords. Examples of the decoding performance are shown in **Error! Reference source not found.** where a single thread containing 10 random messages was transmitted and the process repeated 300 times. Synchronisation was examined using 4 methods; inherent synchronisation with 1 additional static mark, inherent synchronisation with 6 additional static marks, synchronisation from an appended sequence and a combination of a sequence and inherent synchronisation, which applies the inherent synchronisation if no sequence correlation can be detected. Clearly from this we can see that the sequence synchronisation regularly fails to decode where the twitching mirror removes part of the sequence. The use of the inherent synchronisation is much more robust, a situation that is enhanced by adding additional static marks. The combination of sequence and inherent with static appears the most robust. All examples show the system loses some messages on occasion, arising from a combination of the gap removing several principle marks and a small number of inherent marks. Clearly it would be possible to add significantly more static marks to improve the robustness further. Nevertheless this test clearly shows that the use of inherent synchronisation significantly improves the robustness.

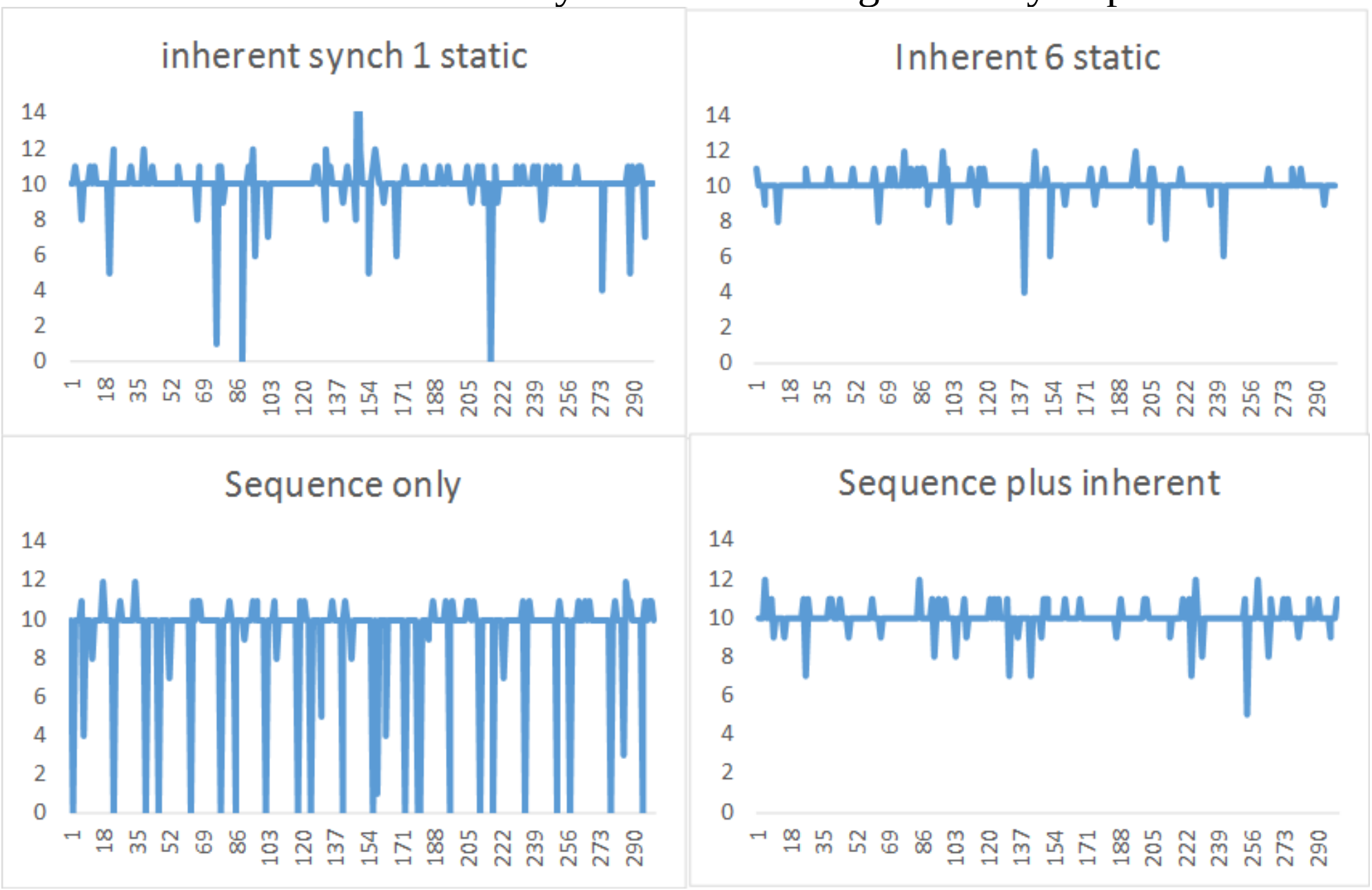

**Figure 7. Plots showing the number of decoded messages for various methods of synchronisation, as a galvonometer mirror introduces burst errors. Each plot is the outcome of 300 decodings using a single thread. Top left uses inherent synchronisation with one static mark included. Top right has 6 static marks. Bottom left uses only a 16-bit correlation sequence for synchronisation. Bottom right combines a sequence with inherent synchronisation.**

Whilst the decoding performance is less than ideal this is the first example of concurrent coding being used with an optical signal and the first example of inherent synchronization ever being implemented. There is nonetheless room for improvement.

### *Multiple Threads*

Having now established a method of synchronisation we can demonstrate the use of multiple threads. Codewords generated from different hash functions were sent to four spatially separated LEDS. With each LED transmitting a separate signal that is overlaid and indistinguishable from the other LEDS, this system behaves as if there are individual users or channels communicating simultaneously with a single receiver. In this case all the threads and streams are synchronised to produce a single codeword but this need not be the case in general. Decoding proceeds by correlating the principle marks (including multiple static marks) for the first hash/thread to establish synchronisation and mark position correction. Then multiple stages of decoding are performed to decode each thread. In this case 5 random messages per thread were encoded into 4 threads and the decoding results are seen in **Error! Reference source not found.**.
For each thread the average number of decoded messages was 5.16, meaning an average of 0.16 hallucinations per decoding. With no burst errors introduced there is no loss of data. This represents the first demonstration of the use of concurrent coding over an optical channel and the first demonstration of multiple overlaid threads or users with concurrent coding.

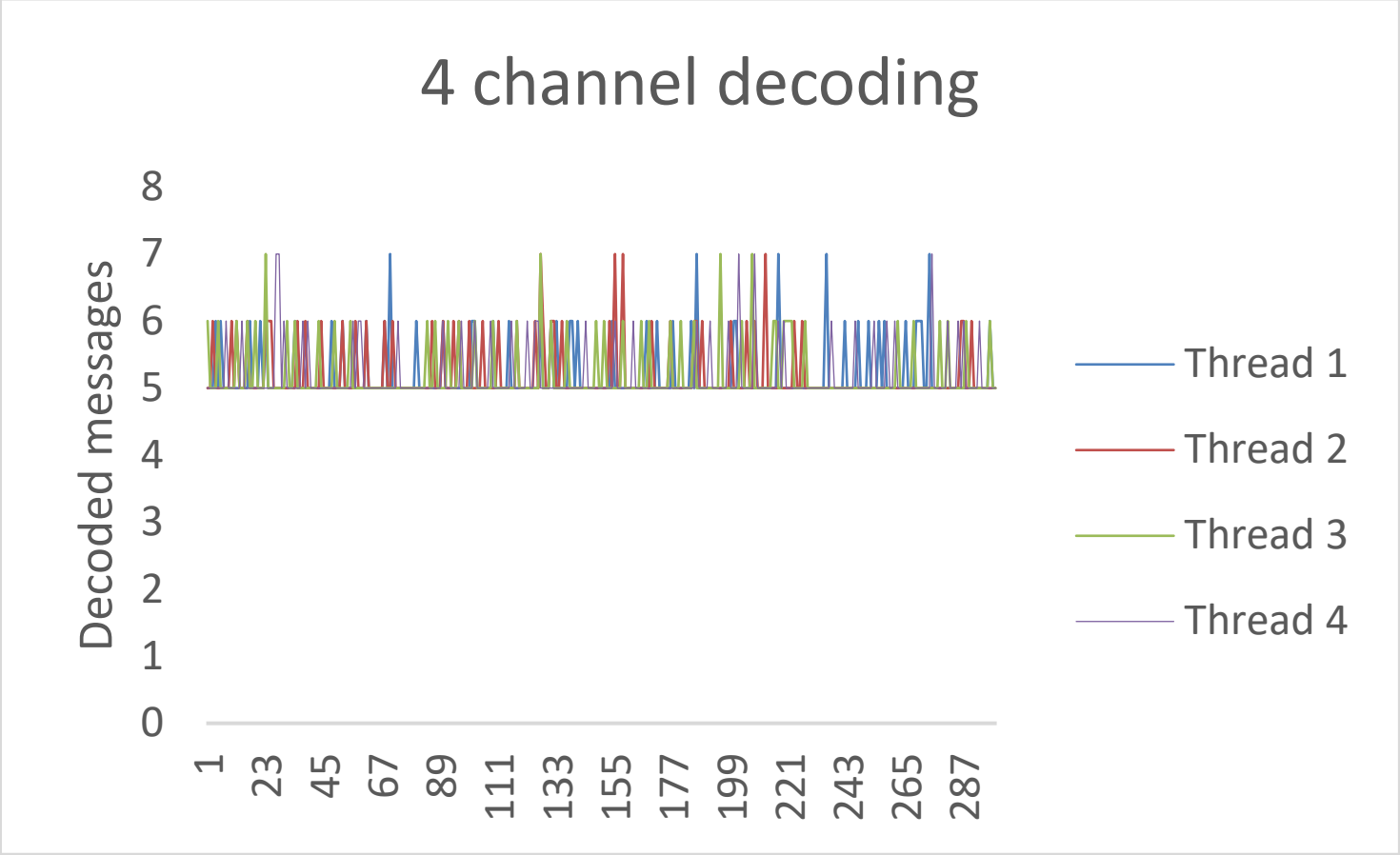

**Figure 8. The number of decoded messages from 300 hundred decodings with 4 simultaneous threads per codeword. 5 messages per thread were transmitted.**

## Conclusions and Discussion

Concurrent coding is a new and novel approach to encoding information which can offer benefits of robustness of data recovery and simplicity of implementation. The use of indelible marks with concurrent coding aligns very well with the OOK modulation which is easily used with free space optical techniques. Concurrent coding offers an alternative method of encoding to protect against both random noise and burst errors. Burst errors are a particular issue for free space communications and concurrent codes have the potential to perform better than interleaving in the recovery of information when burst errors are present. Thus concurrent coding is a potentially interesting tool to employ, particularly as the use of FSO comms is gaining significant interest. In this work characteristics of concurrent coding that will prove useful in future implementations of FSO comms have been investigated. In particular it has been demonstrated that concurrent coding possesses an inherent synchronization structure that allows codeword transmissions to be sent in isolation and reliably decoded with no inclusions or preamble transmissions. Inherent synchronization will help with security concerns (in this and other implementations, such as RF) because the first detected mark is not the start of the codeword transmission and therefore an eavesdropper must understand the hashing function being used to locate the start of the codeword. This also allows the overcoming of burst errors present in the single codeword. In addition it is shown that by encoding information using different hashing functions, that multiple overlaid codeword transmissions can be successfully and independently decoded. This can be considered as a multi-user access channel or a multi-layer transmission channel. It has been shown that this approach works with an optical channel where 4 LED sources transmit information encoded using different hashing functions and are all overlaid onto a single detector. All 4 channels were successfully decoded and this represents the first demonstration of multi-layer concurrent code communication and the first use of concurrent coding over an optical channel.

It is the robustness of concurrent coding that is of interest to FSO comms, particularly where comms is required from mobile or unstable platforms. Mobile applications for FSO inevitably have unpredictable circumstances that require a level of flexibility and robustness of the system. Maintaining a constant link connection between source and receiver can be a hardware problem tackled by accurate beam pointing systems, but it cannot overcome beam interruptions such as a moving object blocking the beam. This is where the encoding protocol helps. Future applications could see mobile sensors or systems needing to send a burst of data and know that the data will be correctly decoded. This could be in the form of compact FSO systems to perform financial transactions, or authorization. Concurrent coding is a tool that will enable this without the need for pre-synchronisation to delay transmission. In addition the efficiency of concurrent coding can reduce the transmitted energy requirements which is always a benefit to mobile systems and has potential defense benefits by having a low probability of intercept.

Concurrent coding is still at an early stage of development and much more needs to be done, particularly around synchronization and its effects upon decoding quality. But concurrent coding offers and alternative way of thinking about encoding for robustness with benefits worth exploring in future applications.

Acknowledgements.

The author would like to thank Paul Brittan for his helpful support and discussions during this work.

**David Benton** graduated in physics from the University of Birmingham in 1989. He completed his PhD in laser spectroscopy for nuclear physics in 1994 and then conducted postdoctoral research in positron emission tomography and then laser spectroscopy for nuclear physics, all at the University of Birmingham. In 1998, he joined DERA, which became QinetiQ, where he worked on a variety of optical projects. He was the leader of a group building quantum cryptography systems and was involved in a notable 140 km demonstration in the Canary Islands. He became chief scientist for L-3 TRL in 2010, working on photonic processing techniques for RF applications. He is now at Aston University with a variety of interests, including innovative encoding techniques, gas sensing, and laser detection techniques.